\documentclass{PoS}
\pdfoutput=1

\usepackage{amsmath}
\usepackage{bm}

\newcommand{\be}{\begin{equation}}
\newcommand{\ee}{\end{equation}}
\newcommand{\bea}{\begin{eqnarray}}
\newcommand{\eea}{\end{eqnarray}}

\newcommand{\Z}{\mathbb{Z}}

\newcommand{\bmu}{\mbox{$\bm{\mu}$}}

\title{Many-flavor Schwinger model at finite chemical potential}

\ShortTitle{Many-flavor Schwinger model at finite chemical potential}

\author{\speaker{Robert Lohmayer}\thanks{Research supported in part by the NSF under grant numbers 
PHY-0854744 and PHY-1205396}\\
        Department of Physics, Florida International University, Miami, FL 33199, USA\\
        E-mail: \email{robert.lohmayer@gmx.net}}

\author{Rajamani Narayanan\thanks{Research supported in part by the NSF under grant numbers 
PHY-0854744 and PHY-1205396}\\
        Department of Physics, Florida International University, Miami, FL 33199, USA\\
        E-mail: \email{rajamani.narayanan@fiu.edu}}

\abstract{
We study thermodynamic properties of the Schwinger model on a torus with $f$ flavors of massless fermions and flavor-dependent chemical potentials. 
Generalizing the two-flavor case, we present a representation of the partition function in the form of a multidimensional theta function and show that the model exhibits a rich phase structure at zero temperature. The different phases, characterized by certain values of the particle numbers, are separated by first-order phase transitions. We work out the phase structure in detail for three and four fermion flavors and conjecture, based on an exploratory investigation of the five, six, and eight flavor case, that the maximal number of coexisting phases at zero temperature grows exponentially with increasing $f$. 

}

\FullConference{31st International Symposium on Lattice Field Theory - LATTICE 2013\\
		July 29 - August 3, 2013\\
		Mainz, Germany}

\begin{document}

\section{Introduction}

QED in two dimensions is a useful toy model to gain an understanding
of the theory at finite temperature and chemical potential~\cite{Sachs:1993zx,Sachs:1995dm,Sachs:1991en}. In
particular,
the physics at zero temperature is interesting since one can study a
system that can exist in several phases.
The theory at zero temperature is governed by two degrees of freedom,
often referred to as the toron variables in a Hodge decomposition of
the U(1) gauge field on a $l\times \beta$ torus, where $l$ is the
circumference of the spatial circle and $\beta$ is the inverse
temperature. Integrating over the toron fields projects on to a state
with net zero charge~\cite{Gross:1980br} and therefore there is no dependence on a flavor-independent chemical potential~\cite{Narayanan:2012du}. The dependence
on the isospin chemical potential for the two-flavor case was studied
in~\cite{Narayanan:2012qf}
and we extend this result to the case of $f$ flavors in this paper. 
After integrating out the toron variables, the dependence on the
$(f-1)$ traceless\footnote{linear combinations that are invariant under uniform (flavor-independent) shifts} chemical potential variables  
can be written in the
form of a $(2f-2)$-dimensional theta function. 
As the same gauge field (toron
variables, in particular) couples to all flavors, this theta function has a non-trivial Riemann matrix.
The
resulting phase structure at zero temperature is quite intricate since it involves
minimization of a quasi-periodic function over a set of integers.
Here, we summarize the results of \cite{Lohmayer2013}, where we derive the theta-function representation of the partition function and work out in great detail the two-dimensional phase
structure for the three-flavor case and the three-dimensional phase
structure
for the four-flavor case.

\section{Partition function}

Consider $f$-flavored massless QED on a finite torus with spatial
length $l$ and dimensionless temperature $\tau=\frac l\beta$. All flavors have
the same gauge coupling $\frac{e}{l}$ where $e$ is dimensionless.
Let 
\be
\bmu^t = \begin{pmatrix}
\mu_1 & \mu_2 & \cdots  & \mu_f\cr
\end{pmatrix}
\ee
be the flavor-dependent chemical potential vector.
The partition function factorizes into bosonic and toronic parts~\cite{Sachs:1991en,Narayanan:2012qf}, 
$Z(\bmu,\tau,e) = Z_b(\tau,e)
Z_t(\bmu,\tau)$.
As we will only consider ourselves with the physics at non-zero chemical potential, we focus on the toronic part
\begin{align}
Z_t(\bmu,\tau) &=
\int_{-\frac{1}{2}}^{\frac{1}{2}} dh_2
\int_{-\frac{1}{2}}^{\frac{1}{2}} dh_1\, \prod_{i=1}^f g(h_1,h_2,\tau,\mu_i)\,,\\
g(h_1,h_2,\tau,\mu) &=
\sum_{n,m=-\infty}^\infty
\exp \left[ -\pi\tau \left[\left(n+ h_2 -i \frac{\mu}{\tau}\right)^2
+\left(m + h_2 -i \frac{\mu}{\tau}\right)^2\right] +2\pi i h_1 \left(n -m\right)\right]\nonumber
\end{align}
 and perform the integration over the toronic variables, $h_1$ and
$h_2$.

\subsection{Multidimensional theta function}

As derived in~\cite{Lohmayer2013}, the toronic part of the partition function has a representation in the
form of a $(2f-2)$-dimensional theta function:
\be 
Z_t(\bmu,\tau)=\frac{1}{\sqrt{2\tau f}}
\sum_{{\bm n}=-\infty}^\infty 
\exp \left[  -\pi\tau \left({\bm n}^t T^t +\frac{i}{\tau} {\bm s}^t\right) 
\begin{pmatrix}
\bar\Omega & {\bm 0} \cr
{\bm 0} & \bar\Omega\cr
\end{pmatrix}
\left( T {\bm n} + \frac{i}{\tau}{\bm s}\right) \right]\label{maineqn}
\ee
where
${\bm n}$ is a $(2f-2)$-dimensional vector of integers.
The $(2f-2)\times (2f-2)$ transformation matrix $T$ and 
the $(f-1)\times (f-1)$ matrix $\bar\Omega$ are given by
\begin{align} 
T &= 
\begin{pmatrix}
1  & 0 & \cdots & 0 & 0\cr
0  & 1 & \cdots & 0 & 0\cr
0  & 0 & \ddots & 0 & 0\cr
0  & 0 & \cdots & 1 & 0\cr
-1  & -1 & \cdots & -1 & f\cr
\end{pmatrix}\,,\qquad\qquad
\bar\Omega = 
\begin{pmatrix}
1 - \frac{1}{f} & -\frac{1}{f} & \cdots & -\frac{1}{f} \cr
- \frac{1}{f} & 1-\frac{1}{f} & \cdots & -\frac{1}{f}\cr
\vdots & \vdots & \ddots & \vdots \cr
- \frac{1}{f} & -\frac{1}{f} & \cdots & 1-\frac{1}{f}\cr
\end{pmatrix}\,.
\end{align}
 The dependence on the chemical potentials comes from
\be\label{eq:s}
{\bm s}^t = 
\begin{pmatrix}
\bar \mu_2 &
\bar \mu_3 &
\cdots &
\bar \mu_f &
-\bar \mu_2 &
-\bar \mu_3 &
\cdots &
-\bar \mu_f \cr
\end{pmatrix}
\ee
where we have separated the chemical potentials into a flavor-independent component
$\bar \mu_1 = \sum_{i=1}^f \mu_i$
 and $(f-1)$ traceless components $\bar \mu_k = \mu_1-\mu_k$ for $2\leq k \leq f$.
 
\subsection{Particle number }

We define particle numbers $N_i$, $\bar N_k$ corresponding to the chemical potentials $\mu_i$, $\bar \mu_k$, resp., as
\begin{align}
  N_i(\bmu,\tau) = \frac{\tau}{4\pi}\frac{\partial}{\partial \mu_i} \ln Z_t(\bmu,\tau)\,,\qquad
   \bar N_k(\bmu,\tau) = N_1(\bmu,\tau)-N_k(\bmu,\tau) \quad\text{for}\ 2\leq k \leq f\,.
\end{align}
In the infinite-$\tau$ limit, the infinite sums in Eq.~\eqref{maineqn} are dominated by ${\bm n}={\bm 0}$ which results in 
\begin{align}
   \bar N_k(\bmu,\infty) = \bar \mu_k \qquad\text{for}\ 2\leq k \leq f\,.\label{numdeninf}
\end{align}
Since the partition function is independent of $\bar \mu_1$, $\bar N_1(\bmu,\tau)=\sum_{i=1}^f N_i(\bmu,\tau)=0$ for all $\tau$.

\subsection{Zero-temperature limit}\label{sec:lowT}

In order to study the physics at zero temperature ($\tau\to 0$), we
set
\be
\Omega = T^t \begin{pmatrix}
\bar\Omega & {\bm 0} \cr
{\bm 0} & \bar\Omega\cr
\end{pmatrix} T.
\ee
Then we can rewrite (\ref{maineqn}) using
the Poisson summation formula as
\be
Z_t(\bmu,\tau) 
=
\frac{1}{\sqrt{2\tau f}\tau^{f-1}}
\sum_{{\bm k}=-\infty}^\infty 
\exp \left[ -\frac{\pi}{\tau}\left( {\bm k}^t
\Omega^{-1} {\bm k} -2 {\bm k}^t T^{-1}
{\bm s}\right)
\right]
\label{zffinal}
\ee
with
\be
\frac{1}{\Omega} = 
\begin{pmatrix}
2 & 1 & \cdots & 1& 1 & 0 & 0 & \cdots & 0 & 1\cr
1 & 2 & \cdots  & 1& 1 & 0 & 0 & \cdots & 0 & 1\cr
\vdots & \vdots & \ddots & \vdots & \vdots & \vdots & \vdots & \ddots
& \vdots & \vdots\cr
1 & 1 & \cdots & 2 & 1 & 0 & 0 & \cdots 0 & 0 & 1\cr 
1 & 1 & \cdots & 1 & 2 & 0 & 0 & \cdots 0 & 0 & 1\cr 
0 & 0 & \cdots & 0 & 0 & 2 & 1 & \cdots & 1 &1 \cr
0 & 0 & \cdots & 0 & 0 & 1 & 2 & \cdots  &1 &1 \cr
\vdots & \vdots & \ddots & \vdots & \vdots & \vdots & \vdots & \ddots
& \vdots & \vdots \cr
0 & 0 & \cdots & 0 &0 & 1 & 1 & \cdots & 2 & 1 \cr
1 & 1 & \cdots & 1 &1 & 1 & 1 & \cdots & 1 & 2-\frac{2}{f} \cr
\end{pmatrix}\label{oinverse}\,,
\ee
where the block in the upper left corner has dimensions $(f-1)\times(f-1)$ and the second block on the diagonal has dimensions $(f-2)\times(f-2)$\,.

For fixed $\bar \mu_k$, the partition function in the zero-temperature
limit is determined by minimizing the term $ {\bm k}^t
\Omega^{-1} {\bm k} -2 {\bm k}^t T^{-1}
{\bm s}$ in the exponent in Eq.~\eqref{zffinal} over the set of integers ${\bm k} \in \Z^{2f-2}$. 
Assuming in general that the minimum is $M$-fold degenerate, let
$S=\{{\bm k}^{(i)}\}_{i=1,\ldots,M}$, ${\bm
  k}^{(i)}\in \Z^{2f-2}$, label these $M$ minima. Then
\begin{align}
\bar N_j(\bmu,0) &=\frac1{2M} \sum_{i=1}^M \left(\sum_{l=1}^{f-1} k^{(i)}_l - \sum_{l=f}^{2f-3} k^{(i)}_l+k^{(i)}_{j-1} -k^{(i)}_{f+j-2}\right)\,,\qquad 2\leq j\leq f-1\,,\\
\bar N_f(\bmu,0) &=\frac1{2M} \sum_{i=1}^M \left(\sum_{l=1}^{f-1} k^{(i)}_l - \sum_{l=f}^{2f-3} k^{(i)}_l+k^{(i)}_{f-1}\right)\,.
\end{align}
If the minimum is non-degenerate (or if all ${\bm k}^{(i)}$ individually result in the same $\bar N_j(\bmu,0) $'s), the particle numbers  $\bar N_j(\bmu,0) $ assume integer or half-integer values at zero temperature. 
Since ${\bm k}\in \Z^{2f-2}$ and we only have $(f-1)$ independent $\bar
  N_j(\bmu,0) $ (with $\bar N_1(\bmu,\tau)=0$ for all $\tau$), there are in general many possibilities to obtain identical particle numbers from different $\bm k$'s. 
The zero-temperature  phase boundaries in the $(f-1)$-dimensional
space of traceless chemical potentials $\bar \mu_{2,\ldots,f}$ are
determined by those $\bm{\bar \mu}$'s leading to degenerate minima
with different $\bm{\bar  N}$'s (for individual  ${\bm k}^{(i)}$'s). Phases
with different particle numbers  will be separated by first-order phase transitions. 
While it is straightforward\footnote{
It is possible
to perform certain orthogonal changes of variables in the space of
traceless
chemical potentials and obtain expressions equivalent to
(\ref{zffinal})
that are more convenient to deal with numerically when tracing the phase boundaries. Such
equivalent
expressions for the case of $f=3$ and $f=4$ are provided in \cite{Lohmayer2013}.
} to numerically determine the phase boundaries at zero temperature (by numerically searching for the minimum), the resulting phase structure turns out to be quite intricate (see details below).

Consider the system at high temperature with a certain choice of
traceless chemical potentials which results in  average values for the
traceless particle numbers
 equal to the choice as per (\ref{numdeninf}). The system
will show typical thermal fluctuations as one cools the system but the
thermal fluctuations will only die down and produce a uniform
distribution of traceless particle numbers  if the initial choice of
traceless chemical potentials did not lie at a point in the phase boundary. Tuning
the
traceless chemical potentials to lie at a point in the phase boundary
will
result in a system at zero temperature with several co-existing
phases. In other words, the system will exhibit spatial
inhomogeneities.

\subsection{Quasi-periodicity}

Changing variables 
${\bar \mu_{k+1}}'= \bar \mu_{k+1} + m_{k}-\frac f2 m_{f-1}+\sum_{i=1}^{f-1} m_i$ for $1 \leq k \leq f-1$
with $m_i\in \Z$ for all $1\leq i \leq f-1$ and $m_{f-1}f/2 \in \Z$, one can show that the partition function is quasi-periodic, resulting in \cite{Lohmayer2013}
\begin{align}\label{eq:shift-N}
\bar N_{k+1}(\bmu',\tau) = \bar N_{k+1}(\bmu,\tau) + m_k -\frac{f}{2} m_{f-1} +\sum_{i=1}^{f-1} m_i,
\end{align}
which is the same as the shift in $\bar\mu$.

\section{Results}\label{results}

\subsection{Phase structure for \bm{$f=2$}}

We partially reproduce the results of \cite{Narayanan:2012qf} in this
subsection.
From Eq.~\eqref{zffinal} for $f=2$, we obtain
\begin{align}
\bar N_2 = \frac{\sum_{k=-\infty}^\infty k e^{-\frac{\pi}{\tau}\left(k- \bar\mu_2\right)^2}}{\sum_{k=-\infty}^\infty e^{-\frac{\pi}{\tau}\left(k-\bar\mu_2\right)^2}}\,.
\end{align}
The quasi-periodicity under $\bar\mu_2' =
\bar\mu_2 + m_1$ ($m_1\in\Z$) is evident. For small $\tau$, the dominating term in the infinite sum is obtained when $k$ assumes the integer value closest to $\bar\mu_2$. Therefore, $\bar N_2(\bar \mu_2)$ approaches a step function in the zero-temperature limit (for plots, see \cite{Narayanan:2012qf} and \cite{Lohmayer2013}).
At zero temperature, first-order phase transitions occur at all half-integer values of $\bar \mu_2$, separating phases which are characterized by different (integer) values of $\bar N_2$.

If a system at high temperature is described in the path-integral
formalism by fluctuations (as a function of the two Euclidean
spacetime coordinates) of $\bar N_2$ around a half-integer value, the
corresponding system at zero temperature will have two coexisting
phases (fluctuations are amplified when $\tau$ is decreased). On the
other hand, away from the phase boundaries, the system will become
uniform at $\tau=0$ (fluctuations are damped when $\tau$ is
decreased).
For visualizations, see \cite{Lohmayer2013}.

\subsection{Phase structure for \bm{$f=3$}}

\begin{figure}[htb]
\centering
\includegraphics[width=0.4\textwidth]{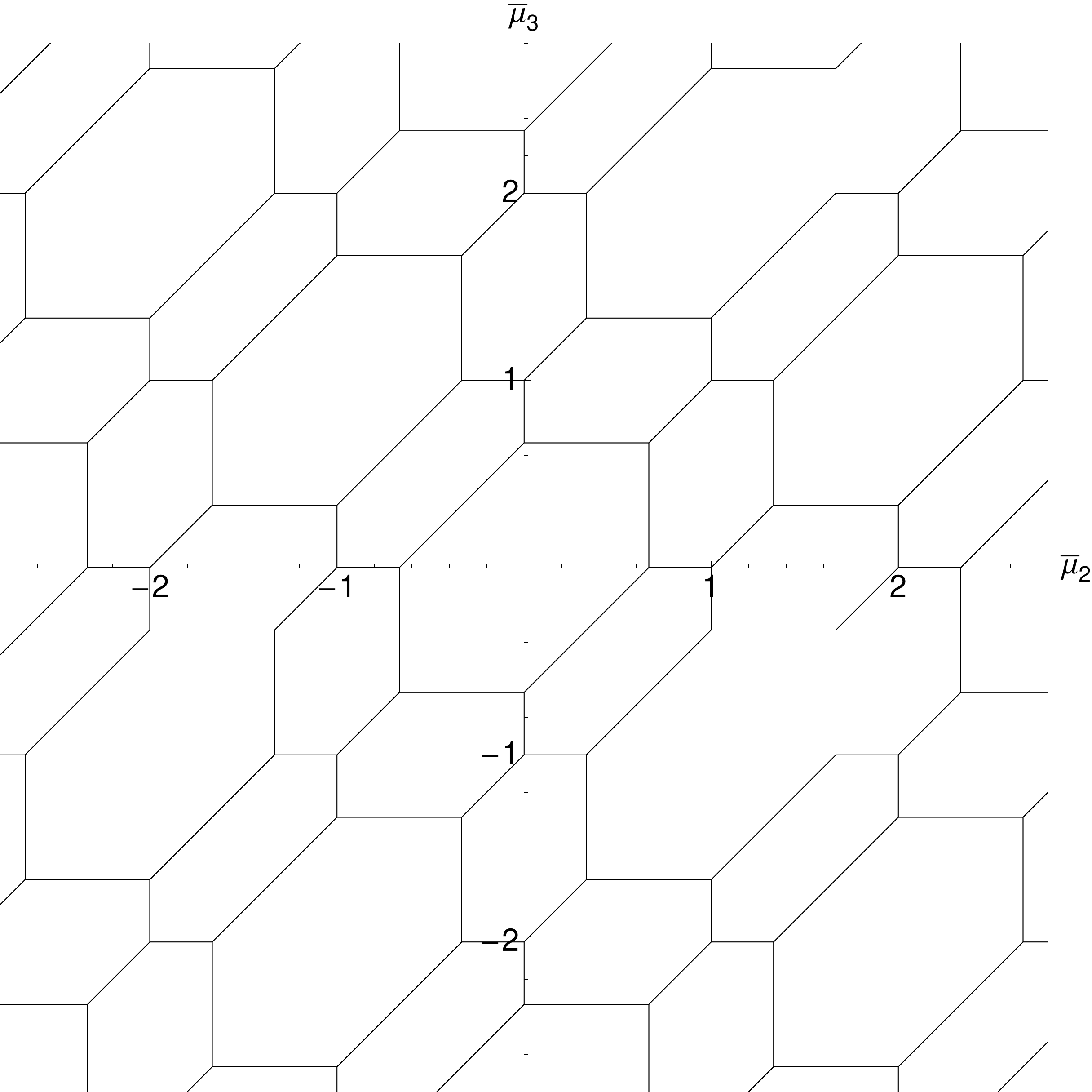}\qquad\qquad
\includegraphics[width=0.4\textwidth]{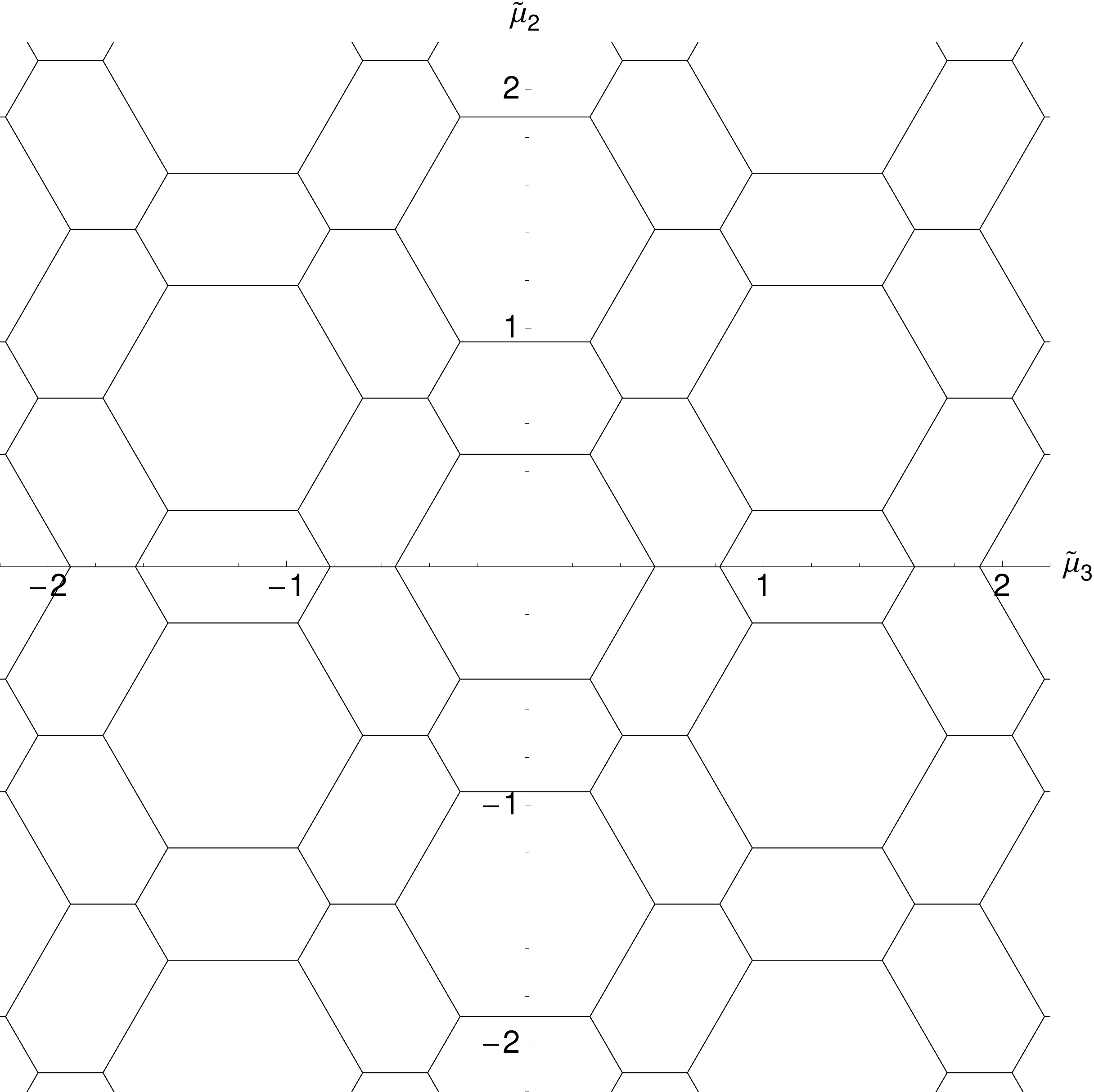}
\caption{Phase boundaries at zero temperature for $f=3$ in the $\bar \mu$ plane (left) and the $\tilde \mu$ plane (right).}
\label{fig:f3-phases}
\end{figure}

We determine the phase boundaries, separating cells with different $(\bar N_2,\bar N_3)$ as described in Sec.~\ref{sec:lowT}. 
As mentioned in Sec.~\ref{sec:lowT}, it is also instructive to use a different coordinate system for the chemical potentials, obtained from $(\mu_1,\mu_2,\mu_3)$ by an orthonormal transformation:  
\begin{align}
\begin{pmatrix}
\tilde \mu_1 \\
\tilde \mu_2 \\
\tilde \mu_3 
\end{pmatrix} 
=
\begin{pmatrix}
\frac 1{\sqrt{3}} & \frac 1{\sqrt{3}} & \frac 1{\sqrt{3}} \\
\frac 1{\sqrt{2}} & -\frac {1}{\sqrt{2}} & 0 \\
\frac 1{\sqrt{6}} & \frac 1{\sqrt{6}} & -\frac 2{\sqrt{6}} \\
\end{pmatrix}
\begin{pmatrix}
\mu_1 \\
\mu_2 \\
\mu_3 
\end{pmatrix} \,.
\end{align} 
 We denote the corresponding particle 
numbers by $\tilde N_2$ and $\tilde N_3$.  An alternative representation of the partition function, which simplifies the determination of vertices in terms of the coordinates $\tilde \mu_i$, is given in \cite{Lohmayer2013}. 
In these coordinates, the phase structure is symmetric under rotations
by $\pi/3$ and composed of two types of hexagonal cells, a central
regular hexagon is surrounded by six smaller non-regular hexagons,
which are identical up to rotations.
Figure~\ref{fig:f3-phases} shows the phase boundaries at zero
temperature in both coordinate systems.

From Eq.~\eqref{eq:shift-N} we see that the boundaries in the $(\bar \mu_2,\bar \mu_3)$ plane are periodic under shifts by integer multiples of $(2,1)$ and $(1,-1)$.

All $\bar \mu$'s inside a given hexagonal cell result in identical $\bar N$ as $\tau\to 0$, given by the coordinates of the center of the cell. For example, $\bar \mu$'s in the central hexagonal cell lead to $\bar N_{2,3}=(0,0)$ at $\tau=0$, the six surrounding cells are characterized by $\bar N_{2,3}=\pm(1,\frac 12)$, $\bar N_{2,3}=\pm(\frac 12,1)$, and $\bar N_{2,3}=\pm(-\frac 12,\frac 12)$. 
Every vertex is common to three cells. The coordinates of the vertices between the central cell and the six surrounding cells are $\pm(\frac 23, \frac 23)$, $\pm(0,\frac 23)$, $\pm(\frac 23,0)$, $\pm(1,1)$, $\pm(0,1)$, $\pm(1,0)$. 

First-order phase transitions occur between neighboring cells with different particle numbers  $\bar N_{2,3}$ at $\tau=0$. At the edges of the hexagonal cells, two phases can coexist, and at the vertices, three phases can coexist at zero temperature. 

In analogy
to the two-flavor case, a
high-temperature system with small fluctuations (as a function of
Euclidean spacetime) of $\bar \mu_{2,3}$ can result in two or three phases
coexisting  or result in a pure state as $\tau\to 0$ depending on the
choice of $\bar\mu_{2,3}$

\subsection{Phase structure for \bm{$f=4$}}

We use Eq.~\eqref{zffinal} to identify the phase structure in the $(\bar \mu_2,\bar \mu_3, \bar \mu_4)$ space, which is divided into three-dimensional cells characterized by identical particle numbers $\bar N_{2,3,4}$ at zero temperature. At the boundaries of these cells, multiple phases can coexist at zero temperature. We find different types of vertices (corners of the cells), where four and six phases can coexist. At all edges, three phases can coexist. 

\begin{figure}[htb]
 \centering
 \includegraphics[width=0.3\textwidth]{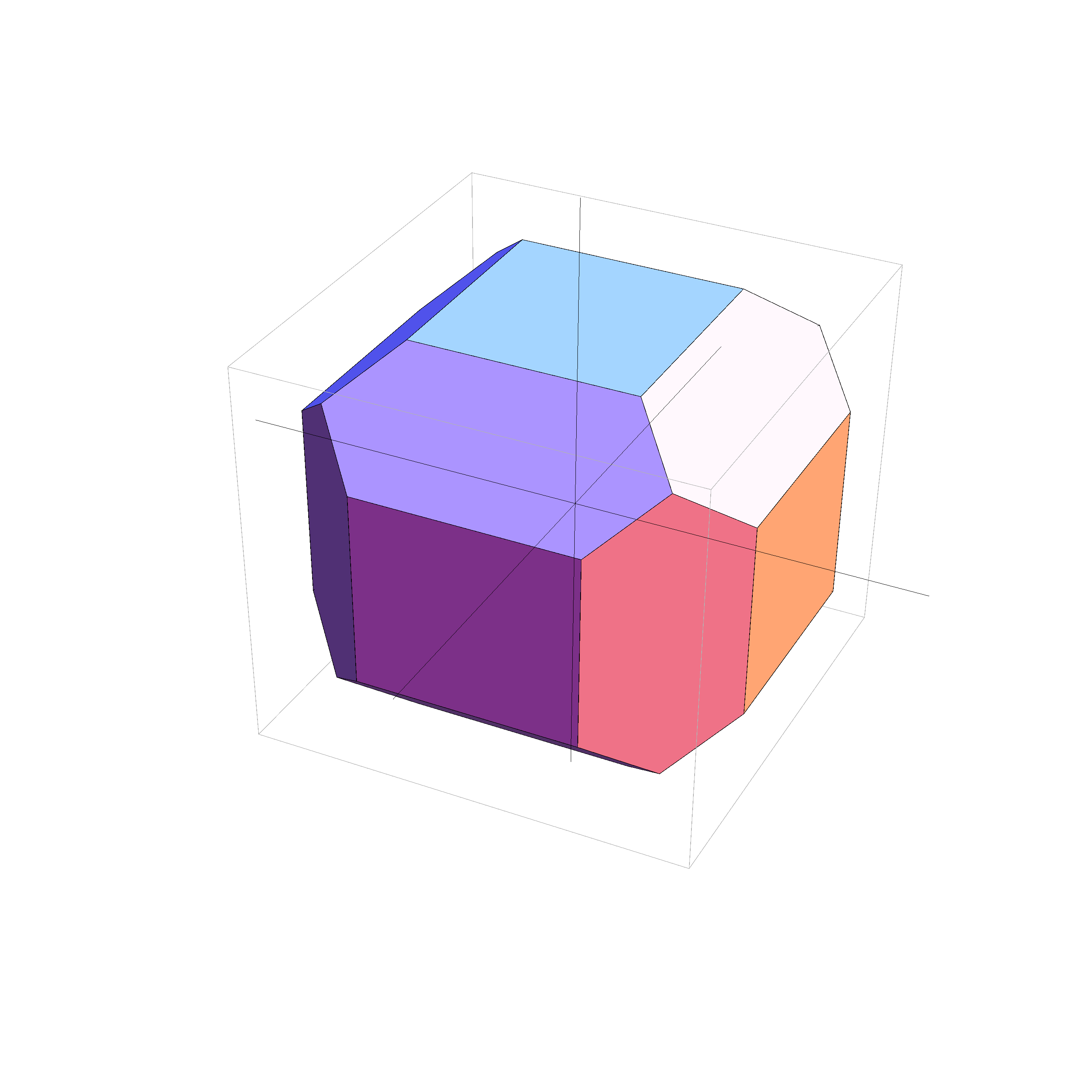}\hfill
 \includegraphics[width=0.3\textwidth]{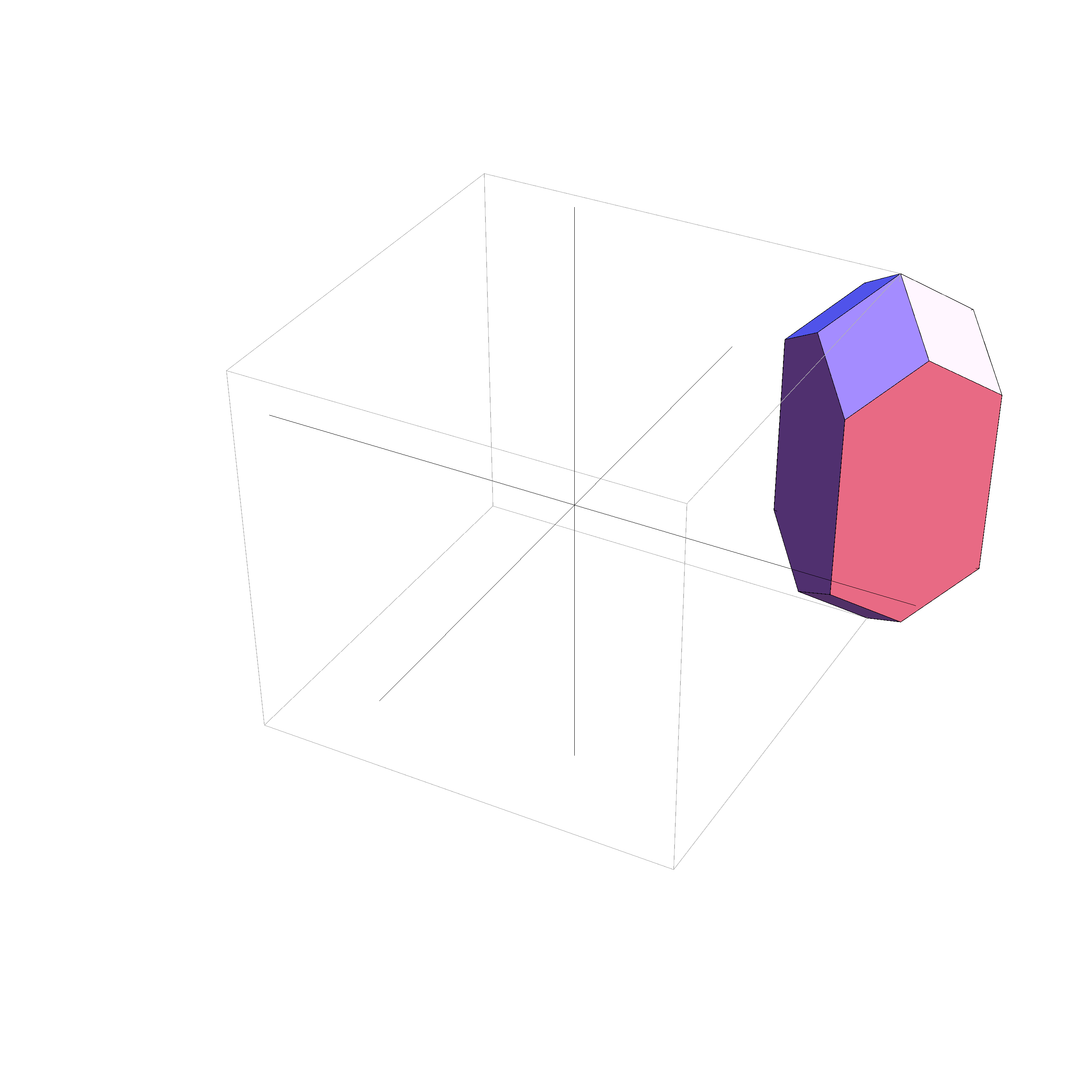}\hfill
 \includegraphics[width=0.3\textwidth]{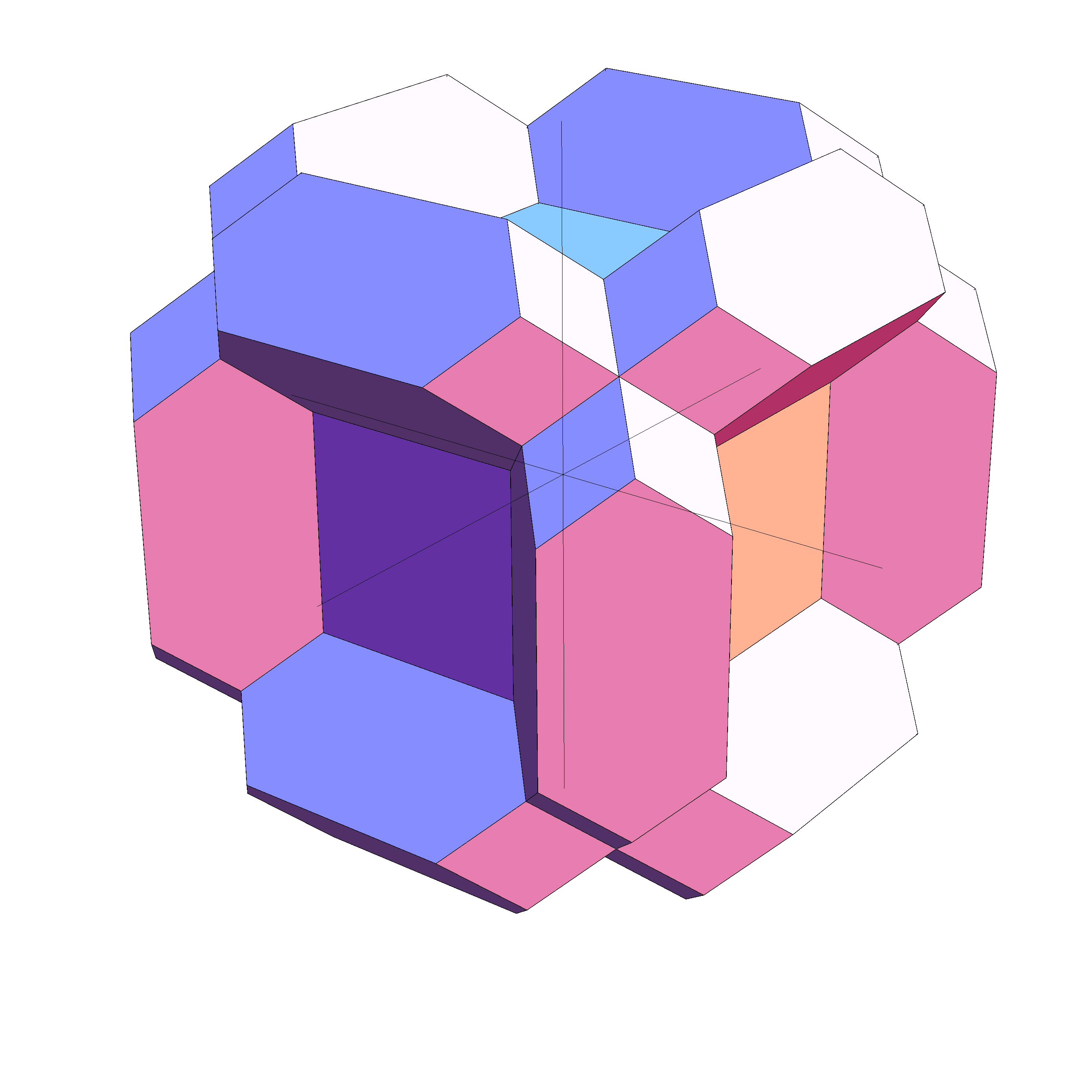}
 \caption{Cells defining the zero-temperature phase structure for $f=4$ in the $\tilde \mu$ coordinates as described in the text. The left figure shows the central ``cubic'' cell, the figure in the center a single ``edge'' cell. The right figure shows the cubic cell together with all 12 attaching edge cells.}
 \label{fig:f4-cells}
 \end{figure}

As in the three flavor case, we observe that the phase structure exhibits higher symmetry in coordinates $\tilde \mu$ which are related to $\mu$ through an orthonormal transformation. A particularly convenient choice for $f=4$ turns out to be given by
\begin{align}\label{eq:trafo-f4}
\begin{pmatrix}
\tilde \mu_1 \\
\tilde \mu_2 \\
\tilde \mu_3 \\
\tilde \mu_4
\end{pmatrix} = 
\frac 12 
\begin{pmatrix} 1 & 1 \\ 1 & -1 \end{pmatrix} \otimes 
\begin{pmatrix} 1 & 1 \\ 1 & -1 \end{pmatrix} 
\begin{pmatrix}
 \mu_1 \\
 \mu_2 \\
 \mu_3 \\
 \mu_4
\end{pmatrix}
\end{align}
since the phase structure then becomes periodic under shifts parallel to the coordinate axes.
The explicit form of an alternative representation of the partition function in these coordinates is given in \cite{Lohmayer2013}.
At zero temperature the $\tilde \mu_{2,3,4}$ space is divided into two types of cells (see Fig.~\ref{fig:f4-cells} for visualizations). We can think of the first type as a cube (centered at the origin, with side lengths 1 and parallel to the coordinate axes) where all the edges have been cut off symmetrically. The original faces are reduced to smaller squares (perpendicular to the coordinate axes) with corners at $\tilde \mu_{2,3,4}=(\pm \frac 12, \pm \frac 14, \pm \frac 14)$ (permutations and sign choices generate the six faces). This determines the coordinates of the remaining 8 corners to be located at $(\pm \frac 38, \pm \frac 38, \pm \frac 38)$. The shift symmetry tells us that these ``cubic'' cells are stacked together face to face. 
The remaining space (around the edges of the original cube) is filled by cells of the second type (in the following referred to as ``edge'' cells), which are identical in shape and are oriented parallel to the three coordinate axes.

This leads to different kinds of vertices (at the corners of the cells described above) where multiple phases can coexist at zero temperature. There are corners which are common points of two cubic and two edge cells (coexistence of 4 phases, for example at $(\pm \frac 12, \pm \frac 14, \pm \frac 14)$), there are corners which are common points of one cubic and three edge cells (coexistence of 4 phases, for example at $(\pm \frac 38, \pm \frac 38, \pm \frac 38)$), and there are corners which are common points of six edge cells (coexistence of six phases, for example at $\tilde \mu_{2,3,4}=(\pm \frac 12,\pm \frac 12,\pm \frac 12)$. Any edge between two of these vertices is common to three cells.

\subsection{Phase structure for \bm{$f>4$}}

One can use the multidimensional theta function to study the phase
structure
when $f>4$ but visualization of the cell structure becomes difficult.
Nevertheless, it is possible to provide examples for the coexistence of many phases. While for $f=5$, we find only up to $5 \choose 2$ coexisting phases,
we find up to $6 \choose 3$ coexisting phases for $f=6$ (for example
at $\bar \mu_{2,\ldots, 6}=(1,\frac{1}{2},0,0,0)$). We also find up to
$8\choose 4$ coexisting phases for $f=8$ (for example at $\bar \mu_{2,\ldots, 8}=(1,1,1,1,1,1,0)$).
This leads us to conjecture that the maximal number of coexisting phases is given by $f \choose{ \lfloor f/2 \rfloor} $, increasing exponentially for large $f$.

\end{document}